\documentclass[10pt,a4paper,oneside,onecolumn]{article}

\usepackage{graphicx}
\usepackage{amssymb}
\begin{document}

\section*{Cosmic signatures in earth's seismic tremor?  \\ 
}

{\bf Francesco Mulargia \\
}

{Dipartimento di Fisica e Astronomia, 
Universit\`a  di Bologna, viale Berti Pichat 8
40127 Bologna, Italy (francesco.mulargia@unibo.it)\\
}

\begin{abstract}
Even in absence of earthquakes, each site on earth experiences continuous elastic vibrations which are mostly traced to the non-linear interactions of sea waves. However, the fine structure of the spectrum at 
mHz frequencies 
shows hundreds of highly significant narrow bandwidth peaks, with a persistence and a coincidence with solar acoustic eigenmodes which are incompatible with any geophysical origin.
The feasibility of a common cosmic origin is evaluated through an estimate of the gravitational wave cross-section of the earth, combined with its elastic response and with the stochastic amplification produced by the interference of the cosmic signal with tremor of oceanic origin.  The measured spectral peaks 
appear  compatible with a gravitational monochromatic illumination at strains $h \gtrsim 10^{-20} $. 
We analize in detail the band around 2.614 mHz, where the binary white dwarf J0651+2844 - which is the second strongest known gravitational stellar source - is expected to emit.
Compatible spectral tremor peaks are found
for both the earth and the sun, but their amplitude is 3 o.m. larger than independent estimates, so that a gravitational source attribution 
would call for a variety of unknown non-luminous sources
with definite mass-distance ratios.

\end{abstract}

{\it Keywords: }
gravitational wave detectors and experiments; noise; seismicity; surface waves and free oscillations;
04.80.Nn, 05.40.Ca, 91.30.Dk, 91.30.Fn


\section*{Introduction}
\begin{figure}
\centering
\includegraphics[width=6in]{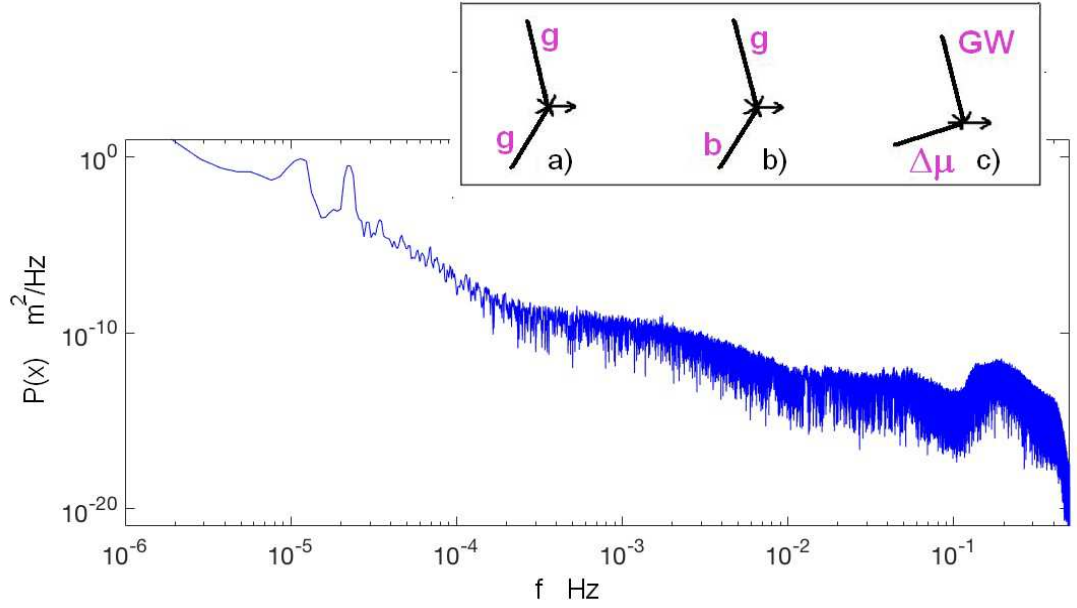}
\caption{ The power spectral density of the vertical component of seismic tremor in terms of displacement $P(x)$ at Obninsk, Russia (OBN)  seismic station in the Jan 1 - Feb 19 2013 interval 
analised with a high resolution multitaper technique.
The inset shows the Feynman diagrams for the three interactions at the origin of seismic noise, where {\bf g} stands for ocean wave, {\bf b} for ocean bottom, 
GW
for gravitational wave and $
{\Delta \mu}$ for rigidity jump.  
}
\label{OBNallFeynman}
\end{figure}

Seismic tremor, i.e., the background motion of the earth in absence of earthquakes, is an 
elastic "noise" wavefield present always and everywhere,  
extending over more than seven frequency decades from $ < 10^{-6}$ Hz to the acoustic band. 
It consists of the stochastic interference of the elastic waves produced by a variety of sources, both natural and anthropic.
 Its amplitude varies largely according to site, time of the day and season,  but its power spectral pattern 
%
remains unchanged, with a strong monotonical decrease with frequency  (in
 displacement), interrupted by a "hump" 
around $\ 200$  mHz  and a lesser one around 60 mHz  (Fig. \ref{OBNallFeynman}).
Seismic tremor is traditionally attributed to oceanic waves,
and in particular to two nonlinear interactions -- the shallow sea wave-sloping bottom interaction and the deep ocean wave-wave interaction -- which, by displacement continuity, excite seismic waves in the solid ocean bottom
[14, 12]. Numerical models  applied to a realistic global bathymetry confirm that these two mechanisms provide an adequate approximation to the general spectral pattern
[24].

\section*{The earth's continuous tremor}
However, the recordings of seismographic and gravimetric stations 
reveal a fine structure with many 
  narrow 
highly significant spectral peaks, 
particularly in the range 1-10 mHz and with large annual cycles in amplitude[18]. 
These go under the name of earth's \em  hum\em, and are commonly identified with the elastic response - i.e., the earth's normal modes - to the continuous broadband stochastic excitation from infragravity sea waves[24]. 
Such an explanation leads to several inconsistencies: 

%
%
%

1) most spectral peaks occur at frequencies within 2\% -  but not in exact coincidence - with low order earth spherical and toroidal eigenmodes[23];  

2) some peaks are extraneous to earth elastic  eigenmodes, but are in surprising coincidence (beyond chance at $6 \sigma$) with solar acoustic eigenmodes[23]; 

3) atmospheric turbulence is not a tenable alternative mechanism, since its persistence is much too short[24];

4) an instrumental artefact origin, produced by magnetic  spurious sensitivity of the seismometers, is implausible[23];

5) several peaks show diurnal and semidiurnal cycles[9], inconsistent with any geophysical origin[24]; 

6) last, but by no means least,  multitaper high resolution spectral analysis reveals that most peaks have a very narrow relative bandwidth $R_B \simeq \Delta \omega / \omega$, standing for quality factors $Q \gg 300$[23]; this is
incompatible with a broadband noise response origin, since it would 
primarily excite (cf. e.g., [16])  
the spheroidal and toroidal modes of the Earth with their related $Q  \lesssim 300 $[8]; the same bandwidth incompatibility  \em a fortiori \em applies
to the broadband noise excitation of atmospheric and oceanic eigenmodes, for which $Q \sim 1$[5].

The above inconsistencies lead to attribute the narrow  spectral tremor peaks 
to their source, which must be monochromatic and with a high $Q$ factor. Since no credible geophysical candidate exists, a source external to the earth must be hypothesized, and  
one proposition along this line was advanced, 
identifying them with
the seismo-magnetic excitation of the earth by the solar acoustic eigenmodes[23].  While the still unspecified acoustic-magneto-elastic coupling at its basis is worth further investigation, we explore here an alternative non-terrestrial origin, i.e.,  that the narrow tremor peaks are produced on both the earth and the sun by the 
 elastic excitation from cosmic gravitational wave monochromatic sources. 
Note how under this perspective the gravitational signature is not \em a signal  above 
noise\em, as it is commonly assumed[1], 
but rather \em a constituent of 
 seismic noise  \em itself (cf. the Feynman diagrams in Fig.\ref{OBNallFeynman}).

\section*{Tremor gravitational signatures}
Choosing a rectilinear coordinate system, as justified by the relative smallness of the earth's mass, and assuming local elastic isotropy,   gravitational waves (from now on GW) induce in solid bodies a displacement $\eta$ [7, 17]. 

\begin{equation}
\frac{\partial}{\partial t} \big( \rho \frac{\partial \eta_i }{\partial t} \big) = \frac{\partial}{\partial x_j} \big[\lambda  \delta_{ij} \epsilon_{kk} + 2 \frac{\partial \mu}{\partial x_j} (\epsilon_{ij} + \frac{1}{2} h_{ij})+ 2 \mu \frac{\partial \epsilon_{ij}}{\partial x_j} \big]\label{boundary}
\end{equation}

\noindent where $\rho$ is the density and $\lambda$, $\mu$ the Lam\'e constants.  
 According to equation (\ref{boundary}), GW produce elastic strains only at rigidity discontinuities, where ${\partial \mu} /{\partial x} \ne 0$.  

The largest elastic body on earth is the earth itself.
Since rigidity is tied to shear wave velocity $v_s$ and density $\rho$ as  $\mu = \rho v^2_s$, there exist two major rigidity discontinuities in the earth, located respectively at the surface
and at the core--mantle boundary, 
where there is a transition from the solid silicate mantle to the liquid iron outer core. 
In terms of rigidity, the largest of these discontinuities is the one at depth, where $\mu$ drops from $2.9 \times 10^{11}$ N m$^{-2}$ to 0, while at the surface it goes from 0 to $2.7 \times 10^{10}$ N m$^{-2}$[8]. However, since the free surface effect provides a factor of $\sim 2$ amplitude increase (cf. e.g., [5]),  
and since we just aim at order of magnitude estimates,
 it will suffice to simply account for 
the total area. 

Let us consider a monochromatic GW cosmic source emitting at frequency $f$. 
The GW induce gravitational elastic waves (from now on GEW) with displacement  $u_j^{GEW}$ dependent on the incidence angle $\theta^{GW}$ to the normal of the surface of the $\mu$ discontinuity

\begin{eqnarray}
u_x & = & \Psi  \sin \theta^{GW} \cos \theta^{GW}  \nonumber \\ 
u_y & = & \Psi  \sin \theta^{GW}    \\
u_z & = & \Psi  \sin^2 \theta^{GW}  \nonumber 
\label{comp}  
\end{eqnarray}

\noindent where $\Psi$ is the gravitational power emitted by the specific cosmic source. 
The interaction of GW with the inner and outer rigidity discontinuities of the earth generates  
$P$ and $S$ waves, the wave-guide interference of which induces, in turn, also Rayleigh and Love waves at the surface and Stoneley waves at the mantle-core interface.
The  major difference between the GEW peaks and those excited by noise lies in 
bandwidth,  since $Q \rightarrow \infty$ for an ideal monochromatic GEW source, while $Q \sim  300$ for broadband seismic noise exciting 
the earth normal modes[8, 5]. 

In fact, such a difference is experimentally apparent between the measured narrow hum peaks - which occur at frequencies not coincident but close to earth normal modes - and the latter (see Fig.\ref{OBN2013Z_4RS}).
Since hum peaks occur within a few percent of the frequency of either a spherical or toroidal earth mode (see Fig.\ref{OBN2013Z_4RS} and cf. [23]) of the earth, which behaves at this frequency as a damped harmonic oscillator with $Q = Q_d \sim 300$, this  implies a response amplification by a factor $q \sim 100$. 
Under a single pulse, this would result in a compatible decay,   accompanied by a progressive spectral broadening and decay by energy transfer to the nearby bands. 
However, under a persistent monochromatic input, equilibrium will be reached after $\sim Q_d$ cycles, and thereafter no further decay will occur, resulting in a
narrow bandwidth peak ruled by the sampling frequency, the record length and the spectral technique adopted (see Fig.\ref{duffing}-top).  

\section*{The gravitational cross section of the earth}
Combining the above response argument with the stationary state of a quadrupole GW source,  the gravitational cross section $\Xi$ at the frequency $f$ of a GW detector with mass $M$ and geometric area $A$  can be written as (cf. [25, 7, 6] 

\begin{equation}
\Xi = \frac{4 \pi^2}{15} \big( \frac{r_g f}{c} \big) q A
\label{cross}
\end{equation}

\noindent where $r_g$ is the Schwarzschild radius, 
$ r_g =
{2 GM}/ {c^2} $,
$G$ is the gravitational constant and 
$q$ is the elastic response at the frequency $f$.  Given the 
radii at core--mantle boundary and at the surface, $A \simeq 6.6 \times 10^{14}$ m$^2$, while
the mass of the detector is that of the mantle + crust, i.e. 
$ 5 \times 10^{24}$ kg, the Schwarzschild radius is $ r_g \sim 7 \times 10^{-3}$ m.
Hence, considering that $ Q_d\approx  300$
[8, 5], i.e., $ q \sim 100$, the gravitational cross section has a quite large extension linearly dependent on frequency: 
for example, at $f=1$ mHz, $\Xi \simeq 4 \times 10^3 $ m$^2$, i.e., the size of football pitch, while at $f=10$ mHz it compares with
Place de la Concorde in Paris.

Following the above discussion on response, let us take the earth as a set of weakly damped harmonic oscillators, which reach thermodynamic equilibrium when the rate of energy dissipation 
is balanced by the absorption of incoming energy. This occurs (cf. Fig.\ref{duffing}-top) in a time 
$ \tau= Q_d/(2 \pi f)$, in which approximately 40\% of the energy is dissipated, while the remaining 60\% is converted into mechanical energy $E$. 
At equilibrium, equipartition assigns a Boltzmann energy $kt$ to each degree of freedom and the latter consists of equal quantities of potential and kinetic energy, while linearity warrants that the excited earth is equivalent to a single oscillator with mass $M$, amplitude $x$ and energy 
$E = \pi^2 f^2 M x^2$.
 After "loading up" for the time $\tau$,
the system will (cf. [6, 20]) absorb GW power $\Psi$ and re-emit GEW of amplitude $x$ as 
\begin{equation}
 \Psi~ \Xi~ \tau = \pi^2 f^2 M x^2
\label{en}
\end{equation}
\noindent where $x$ accounts for the excitation signal $u$ and any amplification this is subject to.  Another amplification beyond the earth's elastic response is  likely to occur.

\section*{Stochastic amplification}
The oceanic wave-wave and wave bottom interactions at the basis of tremor (Fig. \ref{OBNallFeynman}) are highly nonlinear and act upon an excitable system dominated by noise[22]. A particular type of amplification is known to operate in such systems[13]:
\em{stochastic resonance}\em. This can be summarized as the statistical facilitation of the transition to a higher energy state 
by the addition of random noise
[11]. 
 Stochastic resonance occurs when the escape time from a potential well - the Kramers escape time - matches on average the periodicity of a weak forcing, which acts then as a trigger for the transition to another nearby potential minimum, resulting in an amplification of the forcing itself.   Formally amenable to a Fokker-Planck equation, stochastic resonance  is observed in a variety of systems, ranging from mammalian brain neural excitation, bistable ring lasers, semiconductor devices, chemical reactions,  mechanoreceptor cells, etc., 
and can easily lead to $\gtrsim 30 $ dB amplifications[11]. 
An example of this effect, relative to a Duffing nonlinear oscillator, 
 is shown in Fig.(\ref{duffing})-bottom, illustrating how a small monochromatic excitation can be strongly amplified by adding substantial amounts of "tuned" noise. 
 
In the interaction of oceanic waves, stochastic resonance is likely to operate 
 promoting the \em ensemble \em excitation of the system into a local higher energy state - i.e., a seiche -
with transitions stochastically sympathetic with the forcing signal. This occurs at different geographic scales, and it is observed in harbours[19] as well as in oceans[2].
Consistently with the typical values observed for this effect[19], we assume that a further seasonally variable amplification $K$,  up to $ \sim 10^2$, is provided by the interaction with oceanic noise, which acts as a modulator. 

Summing up, the measured 
tremor monochromatic displacement $x$ at the frequency $f_0$ is the original GEW signal $u$ amplified as
\begin{equation}
x =  u   q   K   
\label{ampl}
\end{equation}

\noindent so that, from equation (\ref{en}), the gravitational power flux taking into account the earth's cross section results  

\begin{equation}
\Psi  =  \frac{15 c^3}{ G} \frac {(fx)^2} { A q^2}= \frac{15 c^3}{ G} \frac {(f u K)^2} { A}   
\label{ene}
\end{equation}

\section*{GEW Estimates} 
Let us now estimate the GEW by evaluating the GW power flux required to generate a tremor peak like the ones observed, which in the 1-10 mHz band have power spectral amplitudes from   
$P(x) \sim 10^{-11}  $ m$^2$/Hz to $P(x)\simeq 10^{-14} $ m$^2$/Hz  (see Fig.\ref{OBNallsplit1}). 
The  measured  displacement amplitudes at $f_0$ can then be written 
as (cf. [17])

\begin{equation}
x =  \sqrt {2 P(x) S_{BW}}
\label{vibr}
\end{equation}

\noindent  where 
  $S_{BW} = f_u - f_l$ is the spectral bandwidth,  and $f_u$, $f_l$ respectively the upper and lower corner frequencies around $f_0$.
Now, $S_{BW}$ and is primarily a function of the source, i.e., of its monochromatic character and persistence, but also of the sampling rate and of the techniques used in the spectral analysis. Since after adequate corrections  most cosmic monochromatic sources are stable in excess of $\sim 10^{-8}$ s, GPS accuracy together with record length and the specific spectral technique adopted become the main limiting factors for spectral bandwidth[17]. 
   
Hundreds of tremor narrow spectral peaks, many of which in correspondence with solar acoustic modes,  are found at a statistical confidence level $> 99$\% in the frequency band below 10 mHz 
[18, 23].
Each of these peaks, identified in both gravimetric and seismic recordings through high resolution multitaper spectral spectral techniques,  can be hypothesized as originated by GEW. 


As a specific example, let us consider in detail the frequency band around 2.614 mHz,
which is interesting because it coincides with the 
 supposed GW emission of the binary white dwarf J0651+2844, presently  the second strongest supposed stellar GW emitter.
A narrow spectral peak is apparent at this frequency in the waveforms  of the Black Forest Observatory, Schiltach, Germany (BFO) seismic station in the January-July 2004 period (see Fig.4 of  [23]). 
  Chosing at random among seismic stations distant from the ocean and known for low environmental noise, we consider  51 days of continuous seismic recordings at Obninsk (OBN), Russia,  
  and Tamanrasset (TAM), Algeria,  starting January 1 2013 at 00.00.00.  A winter period is chosen to have the largest amplitude of the narrow spectral peaks for stations located in the northern hemisphere[18]. Only the vertical component is considered in order to avoid the complication of the response to the earth gravity field induced by tilt on the horizontal components[21].  

\begin{figure}
\includegraphics[width=\textwidth]{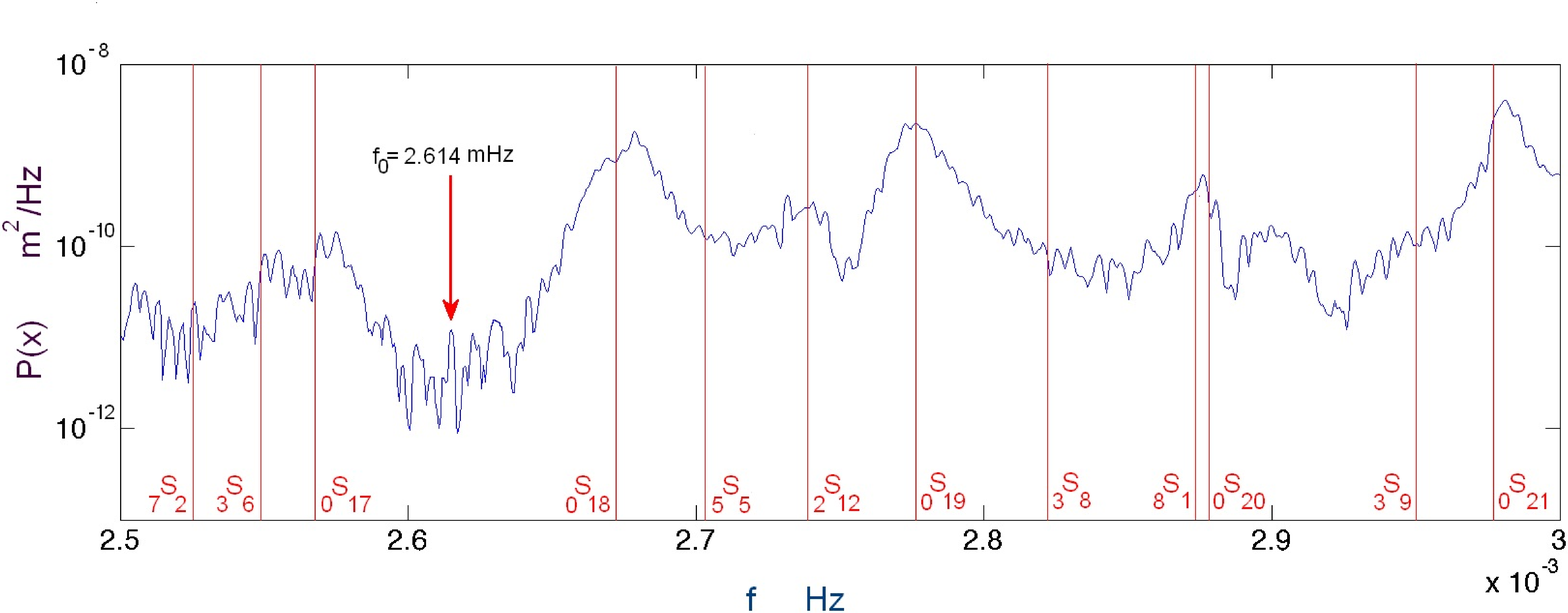}

\caption { The same as in Fig.(\ref{OBNallFeynman}), but 
at the at Tamanrasset, Algeria (TAM) 
seismic station 
and restricted to the frequency band 2.5 - 3.0 mHz, together with the 
earth spherical modes, marked as vertical red lines.
}
\label{OBN2013Z_4RS}
\end{figure}

\begin{figure}
\includegraphics[width=\textwidth]{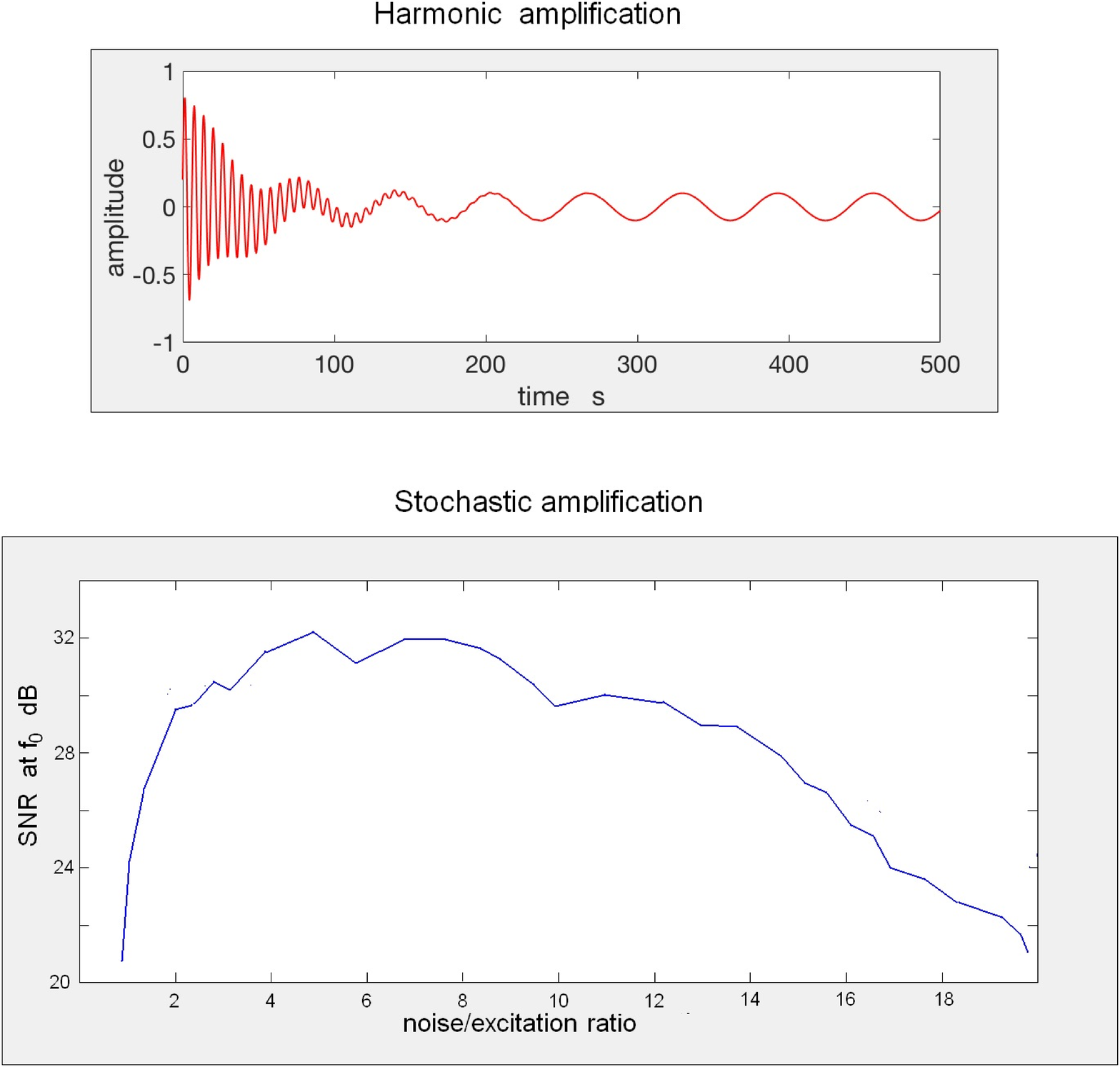}

\caption { Examples of harmonic and stochastic amplification. In the top panel,  a driven and damped  harmonic oscillator (with eigenperiod 5s, amplitude 0.8 and $Q=20$) shows a stable amplification of the forcing signal (with period 60 s and amplitude 0.1) after the initial decay; in the bottom panel, a  Duffing oscillator (see e.g.,[11]) shows the signal to noise ratio (in dB) stochastic amplification versus increasing amounts of added noise. 
}
\label{duffing}

\end{figure}

\begin{figure}
\includegraphics[width=\textwidth]{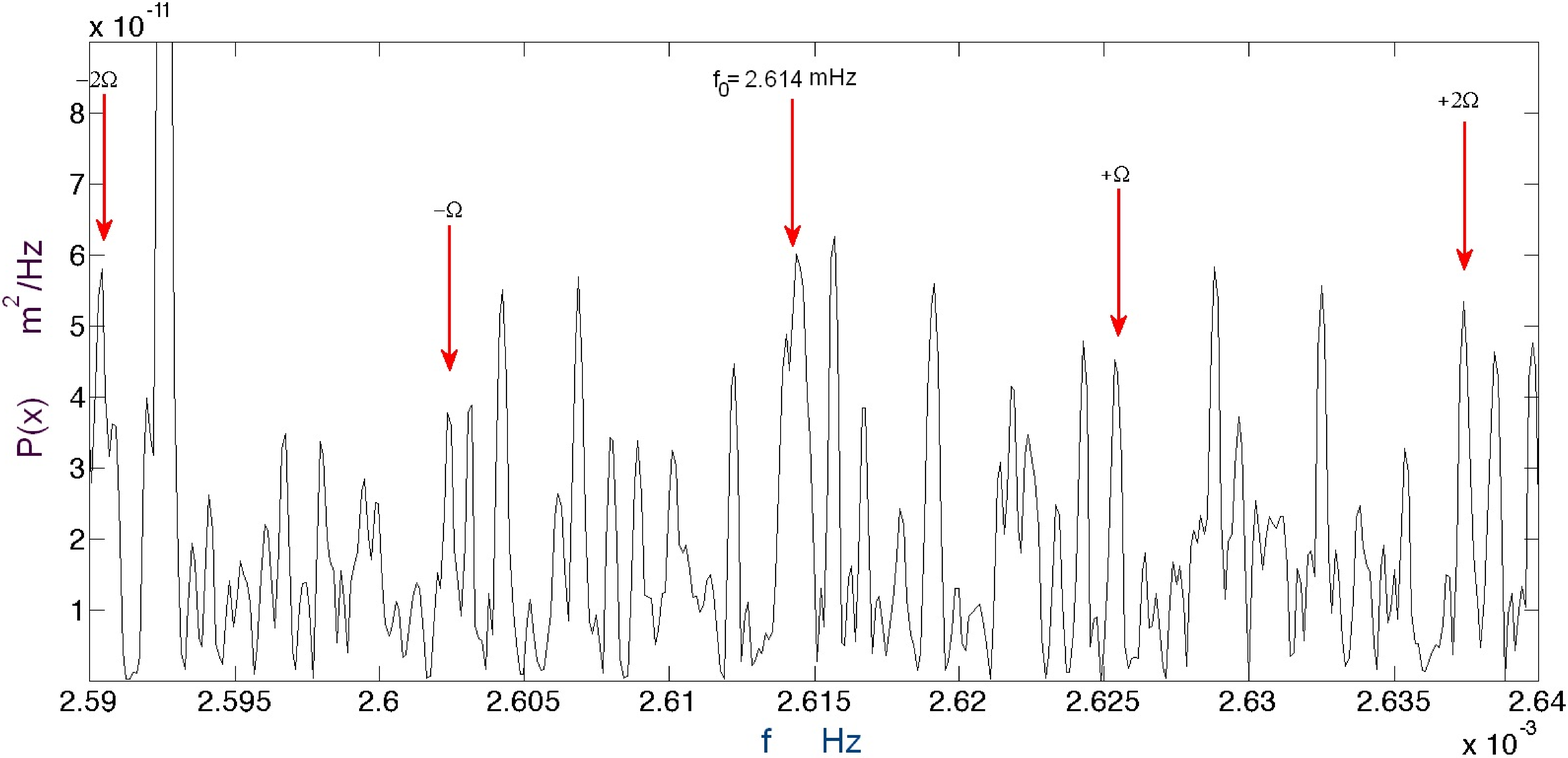}

\caption { The seismic tremor power spectral density  at OBN station, as in Fig. (\ref{OBNallFeynman}), but restricted to the frequency band 2.59 - 2.64 mHz.  The arrows identify the $\pm F, \pm 2 F$ Zeeman splitting produced by earth rotation, with $F=0.01157$ mHz.  Note the linear scales on both axes (cf. [23]).
}
\label{OBNallsplit1}

\end{figure}

Tremor spectral peaks compatible with such a GW illumination are apparent in the recordings (Figs. \ref{OBN2013Z_4RS} and  \ref{OBNallsplit1}), both in correspondence of the central peak at $f=2.614$ mHz, and of the related
 Zeeman splitting[7] into the four singlets $ f \pm  F ,  f \pm 2 F$, produced by the perturbation of the incoming GW signal by earth daily rotation at the frequency $F=0.01157$ mHz. Note that such a splitting 
 is routinely observed right after large earthquakes, which make the whole earth "ring"[5].   
 The spectral pattern shows an obvious difference between the narrow peaks and the much broader earth elastic eigenmodes (Fig. \ref{OBN2013Z_4RS}).  While at mHz frequencies the GPS timing accuracy ($\sim 100$ ns) could provide $S_{BW} \ll 10^{-7}$ Hz,  Fig. (\ref{OBNallsplit1}) suggests a spectral bandwidth $\sim 10^{-6}$ Hz, most likely due to the comparatively short record length analized.
 Note that extending record length could tighten bandwidth, but would also increase the peak height, since the signal amplitude has obviously to remain the same.
Entering the latter value with the measured spectral amplitude $P(x) \sim 8 \times 10^{-11}  $ m$^2$/Hz in equation (\ref{vibr}) yields $x \sim 10^{-8}$ m.  
 
 The 2.614 mHz peak occurs at a frequency
within 2\% of the $_0S_{17}$ (at 2.5687 mHz) and $_0S_{18}$ (at 2.6747 mHz) earth spheroidal modes - and of the solar $P_{0,18}$ pressure mode (at 2.629 mHz)[3] - as well as within a few percent of the earth spherical modes $_0S_{17}$, $_0S_{18}$, $_3S_{6}$, $_7S_{2}$, $_5S_{5}$
(see Fig.\ref{OBN2013Z_4RS}), providing a response amplification of the incoming GEW by a factor $q \sim 100$. Since we consider a winter period and stations in the northern hemisphere, a factor of $ K\sim 100 $ stochastic amplification is  assumed, resulting from eq. (\ref{ampl}) in a GEW amplitude $u \sim 10^{-12}$ m. 
 Hence, considering an average seismic velocity of $\sim 10$ km/sec[8],  a gravitational strain $h \sim  10^{-19}$ is obtained over one cycle. This is $\sim 3$ o.m. above the gravitational emission
estimated for J0651[4], suggesting that
an attribution of the 2.614 mHz tremor peak to this $GW$ source seems presently unrealistic.  

Therefore, considering that this source should be among the strongest GW known emitters, the hypothesis of a GEW origin for the narrow tremor spectral peaks would generally call for non-luminous monochromatic gravitational sources. 
From the above arguments and Kepler's 3rd law we may define the set of binary systems of total mass $M_b$ illuminating the earth with a monochromatic GW power flux $\Psi$ at the frequency $f$ from a distance $d$ as
\begin{equation}
\frac {({f M_b})^{10/3}} {d^2} \simeq  \Psi \frac {c^5} {(2 \pi G )^{7/3}}
\label{Phi}
\end{equation}
\noindent which provides the total mass  - distance source ratios compatible with a measured power flux at a given frequency.  For example, a tremor peak at 10 mHz,  where the measured power density at  $S_{BW} \sim 10^{-6} $ Hz is $P(x)\simeq 10^{-14} $ m$^2$/Hz, stands for a GW strain $h\sim 10^{-20}$ and,  according to equation (\ref{ene}), for a GW flux $\Psi \sim 10^{-5}$   Joule/m$^2$s. According to equation \ref{Phi}, this is compatible, for example,
 with a binary system of rotating black holes of total mass $\sim 8$ solar masses, located at $\sim 1000$ parsec  from the earth. Such an object would have a $\sim 10^5$ km orbital radius and output more than 
 $10^{33}$ J/m$^2$ s in GW, hardly giving any other sign of its existence. 
 It would also be compatible with a system 1 o.m. larger in terms of mass at a $\sim 1$ o.m. larger distance, i.e., at the border of our galaxy.
 
A gravitational attribution to the tremor peaks would first of all account for the coincidence with solar peaks, which would have the same  origin, since the estimated gravitational cross section of the sun[15] essentially coincides with the present estimates for the earth.
It would also be consistent with a cyclic annual dependence, since the stochastic amplification by oceanic tremor has a strong seasonality.   
   Finally, 
  for non-polar GW sources it would account for a diurnal and semidiurnal amplitude variation.
In fact, considering tremor recordings at $N\ge 5$ points - and provided that site-related effects of stochastic amplification can be sorted - the five-fold degeneracy of the gravitational tremor (together with its two polarizations) allows in principle to determine the amplitude and the celestial position of each source[10].

\section*{Acknowledgments}
The seismic tremor waveforms are web available thanks to the IRIS - Incorporated Research Institutions for Seismology.  The author is indebted to the colleagues of the Dipartimento di Fisica e Astronomia dell'Universit\'a di Bologna and of the Istituto Nazionale di Astrofisica di Bologna, and in particular to Sasha Kamenshchick, Luca Ciotti, Roberto Casadio, Michele Cicoli, Andrea Comastri and Alberto Sesana for stimulating discussions. This work was performed with a RFO contribution of the Universit\`a di Bologna, Italy.


\section*{References}
 
\smallskip{[1]} 
 B. Abbott, et al. (LIGO Scientific Collaboration and Virgo Collaboration), 
 Observation of gravitational waves from a binary black hole merger, 
Phys. Rev. Lett., 
116  (2016) 061102.

\smallskip {[2]}
P. Bernard and L. Martel, 
A possible origin of 26 s microseisms, 
Phys. Earth Planet. Inter., 63  (1990) 229-231.
 
 \smallskip{[3]}
A.M. Broomhall, et al.,
Definitive sun-as-a-star p-mode frequencies: 23 years of BiSON observations, 
Mon. Not. R. Astron. Soc.,  100 (2009) 396-400.

\smallskip{[4]}
W. R. Brown, et al.,
A 12 minute orbital period detached white dwarf, 
 Astrophys. J., 737 (2011) L23.
 
 \smallskip{[5]}
F.A. Dahlen and J.T. Tromp, 
Theoretical global seismology,
(Princeton Press, NJ, 1998).

\smallskip{[6]} 
P.C.W. Davies, The search for gravity waves, Cambridge Univ. Press, Cambridge., 1980.

\smallskip{[7]} 
F. J. Dyson, 
Seismic response of the earth to a gravitational wave in the 1-Hz band, 
Astrophys. J.,
156 (1969) 529-540.

\smallskip{[8]}
A. M. Dziewonski  and  D. L Anderson, 
Preliminary reference Earth model, 
Phys. Earth Plan. Interiors, 
25 (1981) 297-356.

\smallskip{[9]}
G. Ekstr\"om,  Time domain analysis of EarthÕs long-period background seismic radiation. 
J. Geophys. Res., 106 (2001) 26483-26494.

\smallskip{[10]}
R. Forward, Multidirectional, multipolarization antennas for scalar amd tensor gravitational radiation, 
Gen. Rel. and Grav., 2 (1971) 149-159.

\smallskip{[11]}
L. Gammaitoni, P. H\"angii, P. Jung, and F. Marchesoni, 
Stochastic resonance, 
Rev. Mod. Phys., 
70 (1998) 223-287.

\smallskip{[12]} 
K. Hasselmann, 
A Statistical Analysis of the Generation of Microseisms, 
Rev. Geophys. Space Phys., 
1 (1963) 177-210.

\smallskip{[13]} 
B. Lindner, J. Garcia-Ojalvo, A. Neiman, and L. Schimansky-Geier, 
Effects of noise in excitable systems, 
Phys. Rep.,  392 (2004) 321-424.

\smallskip{[14]}  
M.S. Longuet-Higgins, 
On the origin of microseseisms
Phil. Trans. Roy. Soc. London, 
243 (1950) 1-35.

\smallskip{[15]} 
I. Lopes and J. Silk, Helioseismology and asteroseismology: looking for gravitational waves in acoustic oscillations,
 Astrophys. J., 794 (2014) 32-39.

 \smallskip {[16]} 
F. Mulargia 
and S. Castellaro, 
Passive imaging in nondiffuse acoustic wavefields, 
Phys. Rev. Lett., 
100  (2008) 218501.  

\smallskip{[17]}
F. Mulargia, and A. Kamenshchik,  Global seismic network as a GW Antenna, 
 Phys. Lett. A,  380 (2016) 1503-1507.

\smallskip{[18]}
K. Nawa, N. Suda, Y. Fukao, T. Sato, Y. Aoyama, and K. Shibuya,  1998.Incessant excitation of the Earth's free oscillations, 
Earth Plan. Space,
50 (1998) 3-18.

\smallskip{[19]}
M. Okihiro, R. T. Guza, and R. J. Seymour,  Excitation of Seiche Observed in a Small Harbor, 
J. Geophys. Res., 98 (1993) 18,201-18,211. 

\smallskip{[20]} 
K. Riles,  
Gravitational Waves: Sources, Detectors and Searches,  arXiv:1209.0667

\smallskip{[21]} 
P.W. Rodgers, The response of the horizontal pendulum seismometer to Rayleigh and Love waves, tilt, and free oscillations of the earth, 
Bull. Seismol. Soc. Am., 58 (1968) 1384-1406.

\smallskip{[22]}
L. W. Schwartz, and J. D. Fenton, 
Strongly nonlinear waves, 
Ann. Rev. Fluid Mech., 14 (1982) 39-60.

\smallskip{[23]}
D. J. Thomson  and F. L. Vernon, III, Unexpected, high-Q, low-frequency peaks in seismic spectra,
 Geophys. J. Int., 
202 (2015) 1690-1710.

\smallskip{[24]}
S.C. Webb, 
The Earth's hum:the excitation of Earth normal modes by ocean waves,
Geophys. J. Int.,
174 (2008) 542-566.
 
\smallskip{[25]}
J. Weber, 
Phys. Rev. Lett., 117 (1960) 306-313.

\newpage

%
%

\end{document}